\PassOptionsToPackage{table}{xcolor}
\documentclass[reprint,aps,prl,floatfix]{revtex4-2}
\usepackage{bm}
\usepackage{amssymb}
\usepackage{amsmath}
\usepackage{graphicx}
\usepackage{rotating}
\usepackage{epsfig}
\usepackage{psfrag}
\usepackage{amsmath}
\usepackage{subfigure}
\usepackage{braket}
\usepackage{tikz}
\usepackage{stmaryrd}
\usepackage{wasysym}
\usepackage{array}
\usepackage{booktabs}
\usepackage{colortbl}
\usepackage[table]{xcolor}

\newcommand{\bea}{\begin{eqnarray}}
\newcommand{\eea}{\end{eqnarray}}
\newcommand{\bpm}{\begin{pmatrix}}
\newcommand{\epm}{\end{pmatrix}}

\usepackage[unicode=true,colorlinks=true,pdfpagemode=UseOutlines,pdfstartview=Fith]{hyperref}
\hypersetup{linkcolor=blue,citecolor=blue,urlcolor=blue}

\begin{document}
\title{Magnetically ordered yet topologically robust phases emerging in concurrent Kitaev spin liquids}

\author{Muhammad Akram$^{1,2}$\textsuperscript{*}}
\author{Aayush Vijayvargia$^1$\textsuperscript{*}}
\author{Hae-Young Kee$^{3,4}$}
\author{Onur Erten$^1$}

\affiliation{$^1$Department of Physics, Arizona State University, Tempe, AZ 85287, USA}
\affiliation{$^2$Department of Physics, Balochistan University of Information Technology, Engineering and Management Sciences (BUITEMS), Quetta 87300, Pakistan}
\affiliation{$^3$Department of Physics, University of Toronto, Toronto, Ontario, Canada M5S 1A7}
\affiliation{$^4$Canadian Institute for Advanced Research, CIFAR Program in Quantum Materials, Toronto, Ontario, Canada, M5G 1M1}

\begin{abstract}
Spin-orbital generalizations of Kitaev model, such as Yao-Lee model, have attracted recent attention due to their enhanced stability of spin liquid phases against perturbations. Motivated by microscopic calculations for the realization of Yao-Lee model showing additional interactions, we study the phase diagram of the Yao-Lee model with added Kitaev and Heisenberg terms. While the plaquette operator is conserved even in the presence of added perturbations, the model becomes no longer exactly solvable. Using perturbation and Majorana mean-field theory, we find magnetic order can arise in the spin sector while the orbital sector remains a liquid for dominant Kitaev interactions, whereas both sectors form liquid phases when Yao-Lee interactions dominate. Additional Heisenberg exchange can enhance or suppress the magnetic order, revealing a rich coexistence of magnetic and topological phases.
\end{abstract}

\maketitle

Quantum spin liquids (QSLs) are characterized by the absence of long-range magnetic order. Due to their underlying topological character, they exhibit exotic emergent properties such as fractionalized quasiparticles\cite{Broholm_Science2020, Balents_Nature2010, Savary_RepProgPhys2016}. The Kitaev model on honeycomb lattice\cite{Kitaev_AnnPhys2006} represents a key theoretical model since it is one of the first models that is exactly solvable with a QSL ground state. Significant progress has been made in recent years towards identifying materials with strong Kitaev-type interactions\cite{jackeli2009mott,Rau_PRL2014,winter2016challenges,Rau2016ARCMP,hermanns2018physics,motome2020hunting,Takayama2021JPSJ,Trebst_PhysRep2022,rousochatzakis2024beyond}, with $\alpha$-RuCl$_3$\cite{plumb2014alpha, Takagi_NatRevPhys2019,Matsuda_arXiv2025} and iridates\cite{jackeli2009mott,singh2012relevance,gegenwart2015kitaev,HwanChun_NatPhys2015,Kitagawa_Nat2018} as notable examples. Yet, an unambiguous confirmation of a Kitaev spin liquid has yet to be achieved.

Although theoretically elegant, the Kitaev model is highly susceptible to perturbations. For example, for an effective spin model with symmetry-allowed terms for $\alpha$-RuCl$_3$, the Kitaev spin liquid occupies only a small portion of the phase diagram\cite{Rau_PRL2014}. This fragility arises in part because most perturbations do not commute with the flux (plaquette) operator, therefore introduce quantum fluctuations that destabilize the QSL ground state.

One approach to addressing the fragility of the Kitaev model is to consider its spin-orbital generalizations\cite{Wu_PRB2009, Yao_PRL2011, Carvalho_PRB2018, Chulliparambil_PRB2020, Natori_PRL2020, Chulliparambil_PRB2021, Seifert_PRL2020, Tsvelik_PRB2022, Nica_npjQM2023, Akram_PRB2023, Vijayvargia_PRR2023, Keskiner_PRB2023, Poliakov_PRB2024, Majumder_PRB2024, Dutta_arXiv2025, Neehus_arXiv2025}. These models have Kugel-Khomskii type interactions\cite{Kugel_SovPhys1982} and their enlarged local Hilbert space allow for a broader class of perturbations that commute with the flux operators and thus preserve the integrability of the model.  One notable example is the Yao-Lee (YL) model\cite{Yao_PRL2011} defined on a honeycomb lattice. The ground state of YL model possesses three flavors of gapless Majorana fermion excitations and the model remains exactly solvable even in the presence of external magnetic field\cite{Chulliparambil_PRB2021}, Dzyaloshinskii-Moriya interaction\cite{Akram_PRB2023} and couplings to conduction electrons\cite{Coleman_PRL2022} or local moments\cite{Keskiner_MatTodayQ, Akram_arXiv2025}.
\begin{figure}[t]
    \centering\includegraphics[width=1\linewidth]{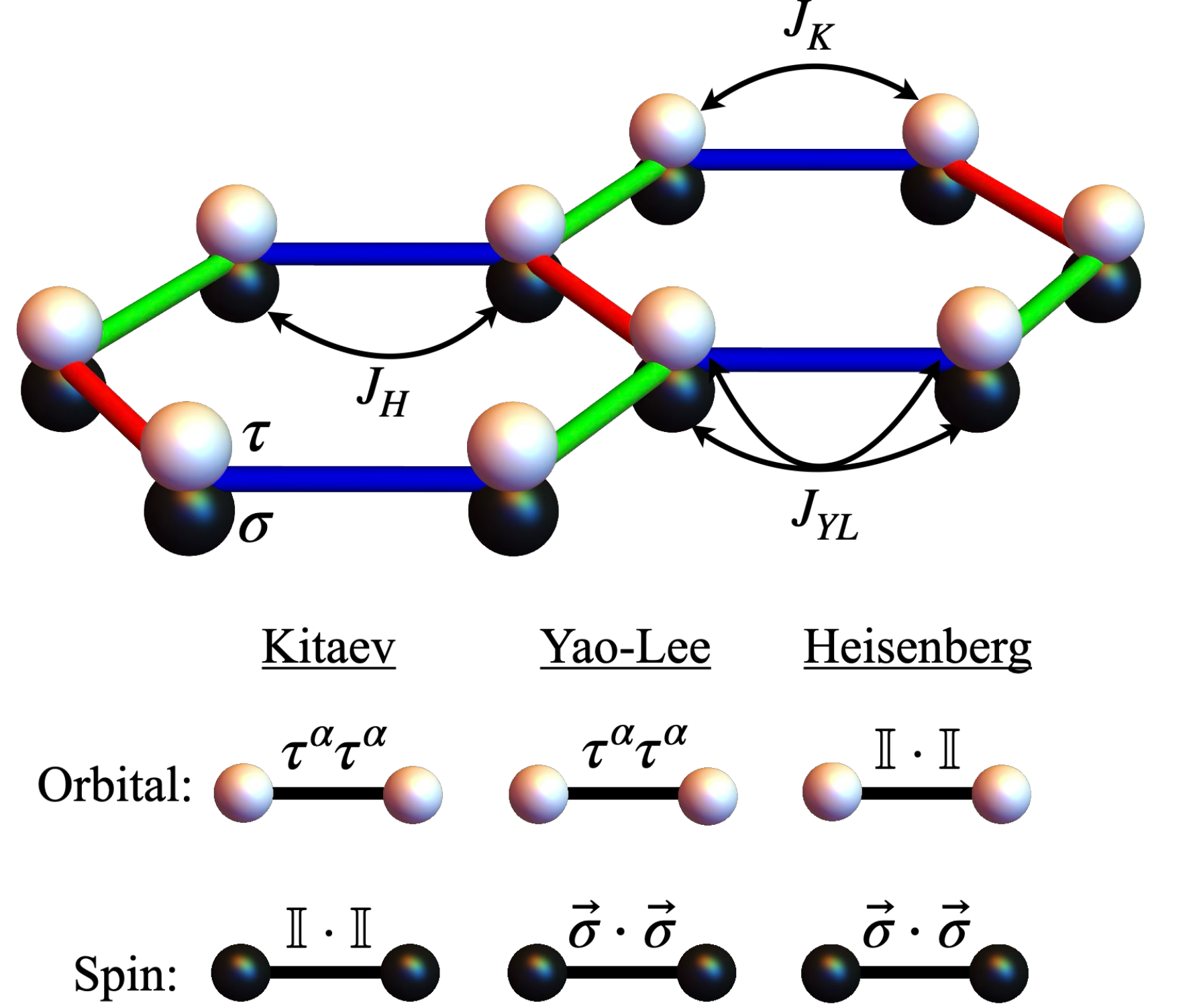}   \caption{Schematic of the model: A section of the honeycomb lattice where the black and white spheres depict the spin ($\sigma$) and orbital ($\tau$) degrees of freedom respectively. The three terms of our full Hamiltonian are pictorially depicted by differentiating their spin and orbital sector. The red, blue and green colors represent $\alpha=1,2,3$ type bonds.}
    \label{Fig:1}
\end{figure}
A recent study\cite{Churchill_npjQM2025} demonstrated that the YL model can be realized in an edge-sharing octahedral environment with partially filled $e_g$ orbitals and strongly spin-orbit coupled ligands. However, beyond the dominant YL interaction, additional terms such as Kitaev and Heisenberg interactions also arise. Motivated by this, we investigate the phase diagram of the YL model in the presence of additional Kitaev and Heisenberg terms via perturbation theory and Majorana mean field theory. Although the plaquette operator of the YL model commutes with both additional terms, the model is no longer exactly solvable in the presence of either term. Our key findings are: (i) For dominant Kitaev interaction, the spin sector exhibits magnetic order—ferromagnetic or antiferromagnetic depending on coupling signs—while the orbital sector remains in a liquid phase. This magnetically-fragmented state displays coexisting local and topological order. (ii) For dominant YL interaction, both spin and orbital sectors retain their liquid character without any symmetry breaking. (iii) The Heisenberg interaction tunes the magnetic order, either enhancing or suppressing it based on its sign.

{\it Microscopic model.} We consider the following Hamiltonian that has YL, Kitaev and Heisenberg interactions, $H=H_{YL}+H_{K}+H_{H}$, as depicted in Fig.~\ref{Fig:1}.
\begin{eqnarray}
H_{YL}&=&J_{YL}\sum_{\langle ij \rangle_{\alpha}}(\boldsymbol{\sigma}_i \cdot \boldsymbol{\sigma}_j)(\tau_i^\alpha \tau_j^\alpha) \label{eq:YL}\\
H_{K}&=&J_K\sum_{\langle ij \rangle_\alpha} \tau_i^\alpha \tau_j^\alpha \label{eq:K} \\
H_{H}&=&J_H \sum_{\langle ij \rangle} \boldsymbol{\sigma}_i \cdot \boldsymbol{\sigma}_j
\end{eqnarray}
Here, $\sigma$ and $\tau$ are two sets of Pauli matrices corresponding to different degrees of freedom (DOF). For the remainder, we will refer $\sigma$ as spin and $\tau$ as orbital DOF; but these notations are interchangeable as demonstrated in Ref.~\citenum{Churchill_npjQM2025}. As such, the local Hilbert space is four-dimensional, spanned by the $S=3/2$ operators or $\Gamma$ matrices. Note that the Kitaev ($H_K$) and Heisenberg term ($H_H$) only act on the orbital and spin DOF respectively while the YL term ($H_{YL}$) acts on both. The plaquette operator, $W_p = \tau_i^x \tau_j^y \tau_k^z \tau_l^x \tau_m^y \tau_m^z$ commutes with the Hamiltonian, $[H,W_p]=0$, and therefore the eigenstates of $H$ can be labeled by the eigenvalues of $W_p = \pm 1$.

\begin{figure}[t]
   \centering
    \includegraphics[width=\linewidth]{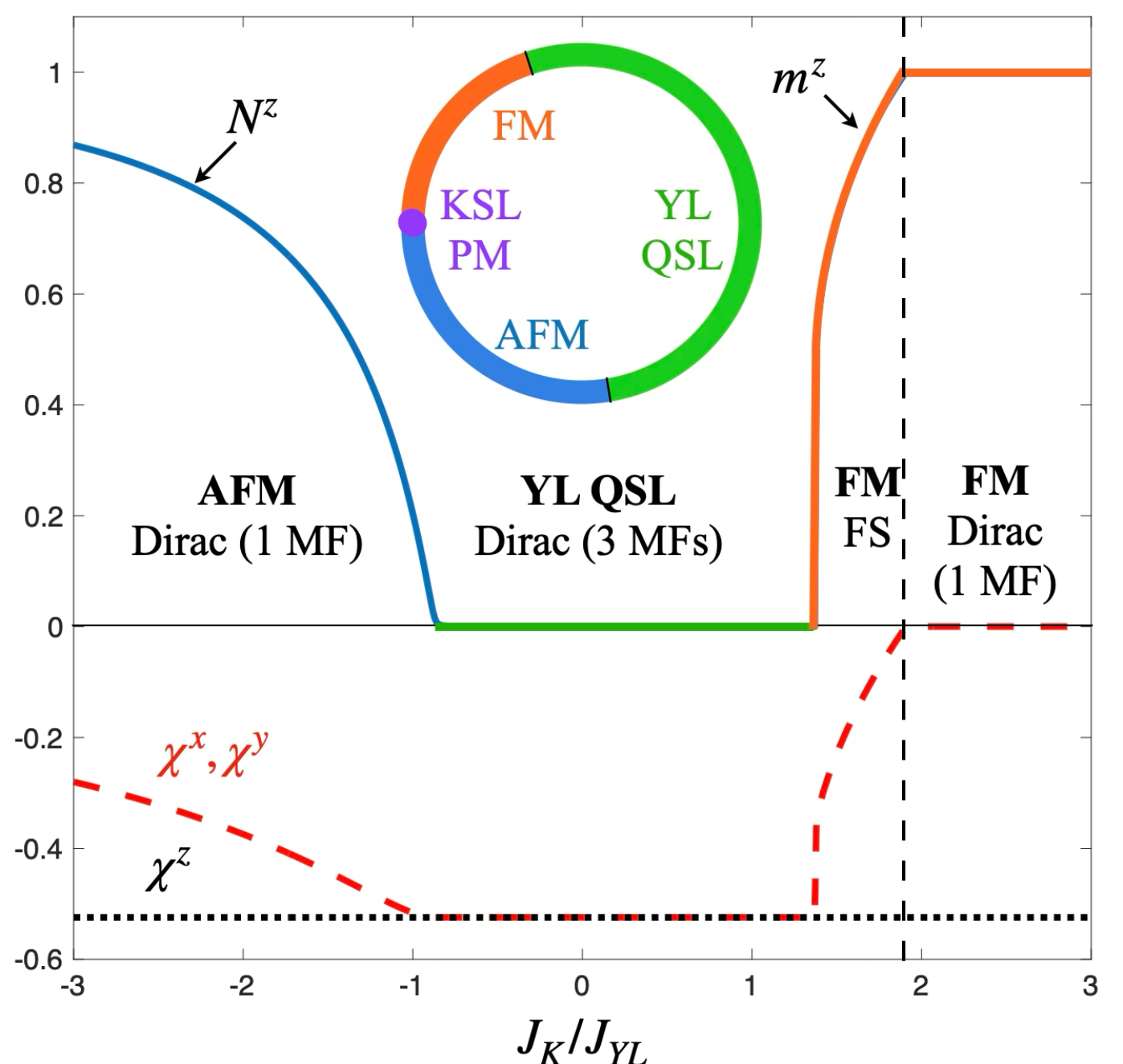}
    \caption{Majorana mean field phase diagram as a function of $J_K/J_{YL}$ ($J_H=0$). For $J_K/J_{YL}<0$, a second order phase transition to an antiferromagnetically ordered phase takes place at $J_K/J_{YL}<-0.85$ which
    gaps two of the three Majorana fermion flavors. For $J_K/J_{YL}>0$, a first order ferromagnetic transition at $J_K/J_{YL}<1.38$ shifts the Dirac cones, giving rise to a Dirac dispersion coexisting with a Fermi surface. A Lifshitz transition at $J_K/J_{YL}<1.91$ gaps the Fermi surface, leaving a single Dirac cone. Inset: The phase diagram on a circle with $\theta=2\arctan (J_K/J_{YL})$. The purple point emphasizes the $\theta=\pi(-\pi)$ point i.e., the infinite $J_K$ limit where the Kitaev spin liquid in $\tau$ sector coexists with a paramagnet (PM) in $\sigma$ sector. }
    \label{Fig:2}
\end{figure}

\noindent
{\it Majorana fermion representation.} It is convenient to reexpress the Hamiltonian by four-dimensional $\Gamma$ matrices as follows, $\Gamma^\alpha = -\sigma^y \otimes \tau^\alpha,~ \Gamma^4 =\sigma^x \otimes \mathbb{I}_2,~ 
\Gamma^5 = -\sigma^z \otimes \mathbb{I}_2$. $\Gamma$ matrices obey the Clifford algebra $\{\Gamma_{i}, \Gamma_j \} = 2\delta_{ij}$. There are five $\Gamma^{\alpha}$ operators and ten $\Gamma^{\alpha \beta}=\frac{i}{2}[\Gamma^{\alpha},\Gamma^{\beta}]$ along with an identity matrix, which forms a complete local Hilbert space. In terms of $\Gamma$ matrices, the Hamiltonian can be written as $H_{YL} = J_{YL} \sum_{\langle ij \rangle_\alpha} (\Gamma_i^\alpha \Gamma_j^\alpha +  \Gamma_i^{\alpha 4}\Gamma_j^{\alpha 4} + \Gamma_i^{\alpha 5}\Gamma_j^{\alpha 5})$; $H_{K} = \frac{J_{K}}{2}\sum_{\langle ij \rangle_\alpha} \epsilon^{\alpha \beta \gamma} \Gamma_i^{\beta\gamma}\Gamma_j^{\beta\gamma}$; $H_H = J_H \sum_{\langle ij \rangle} \Gamma_i^{45}\Gamma_j^{45}+\Gamma_i^{4}\Gamma_j^{4}+ \Gamma_i^{5}\Gamma_j^{5}$. Next, we introduce 6 Majorana fermions per site, $\Gamma_i^\alpha = ib_i^\alpha c_i$ and $\Gamma_i^{\alpha \beta} = i b_i^\alpha b_i^\beta$. In terms of Majorana fermions, the Hamiltonian takes the form,
\begin{eqnarray}
H_{YL}&=&J_{YL} \sum_{\langle ij \rangle_\alpha} u_{ij} (ic_i^xc_j^x+ic_i^yc_j^y +ic_i^zc_j^z) \\
H_{K} &=& -\frac{J_{K}}{2} \sum_{\langle ij \rangle_\alpha} \epsilon^{\alpha \beta \gamma}b_i^\beta b_i^\gamma b_j^\beta b_j^\gamma \\
H_{H}&=&-J_H\sum_{\langle ij \rangle}(c_i^yc_i^zc_j^yc_j^z+c_i^zc_i^xc_j^zc_j^x+c_i^xc_i^yc_j^xc_j^z)
\end{eqnarray}
where $u_{ij} = -i b_i^\alpha b_j^\alpha$. Note that we relabeled $c_i \rightarrow c_i^y,~ b_i^4 \rightarrow c_i^z,~ b_i^5 \rightarrow c_i^x$ to match the notation of previous works\cite{Yao_PRL2011, Natori_PRL2020, Nica_npjQM2023, Akram_PRB2023}.
The plaquette operator can be expressed in terms of the bond operators, $W_p=\prod_p u_{ij}$. The Majorana fermion representation is overcomplete and the physical states obey the constraint $D_i = i b_i^1 b_i^2 b_i^3 c_i^x c_i^y c_i^z = 1$. This constraint can be imposed by a projection operator $\psi_{\rm Phys} = \prod_i \frac{(1+D_i)}{2}\psi$. Note that while the YL model simplifies to bilinears in MFs, the Kitaev model does not. Using the $D$ identity, the Kitaev term can be reexpressed as
\begin{eqnarray}
    H_K = -J_{K} \sum_{\langle ij \rangle_\alpha} iu_{ij} c_i^x c_i^y c_i^z  c_j^x c_j^y c_j^z
    \label{eq:HK}
\end{eqnarray}
Note that an alternative Majorana fermion representation starting with four Majorana fermions each for $\sigma$ and $\tau$ and projecting down to 6 Majorana fermions using constraints leads to the same results\cite{Tsvelik_PRB2022} (see supplemental material for the derivation). 

Before delving into the phase diagram involving multiple terms in the Hamiltonian, we briefly review the limiting cases. In the YL limit ($J_K, J_H = 0$), the ground state is a spin-orbital liquid with three flavors of itinerant Majorana fermions given by $c^x,~c^y$ and $c^z$. In the Kitaev limit ($J_{YL}, J_H=0$), we anticipate having an orbital liquid in $\tau$ sector and a paramagnet in the spin sector. However it is not obvious how eq.~\ref{eq:HK} reduces to this form. Note that the local spin operators can be expressed as $\sigma_i^\alpha = \frac{i}{2} \epsilon^{\alpha \beta\gamma} c_i^\beta c_i^\gamma$. In addition, any $(ic_i^\alpha c_i^\beta)$ commutes with $H_K$ and therefore it is a constant of motion and takes eigenvalues $\pm 1$. These correspond to the paramagnetic spin components and due to the two-fold degeneracy per site, they give rise to a total $2^N$ degeneracy to the ground state. Once any two of the $(c^x,\ c^y,\ c^z)$ Majorana fermions per site are decoupled from $H_K$, the remaining single Majorana fermion corresponds to the delocalized Majorana fermion in the Kitaev model. We remark that $(ic_i^\alpha c_i^\beta)$ does not commute with the Hamiltonian for finite $J_{YL}$ or $J_H$ and consequently $H_K$ does not simplify to bilinears in Majorana fermions in that case.

\begin{figure}[t]
   \centering
    \includegraphics[width=\linewidth]{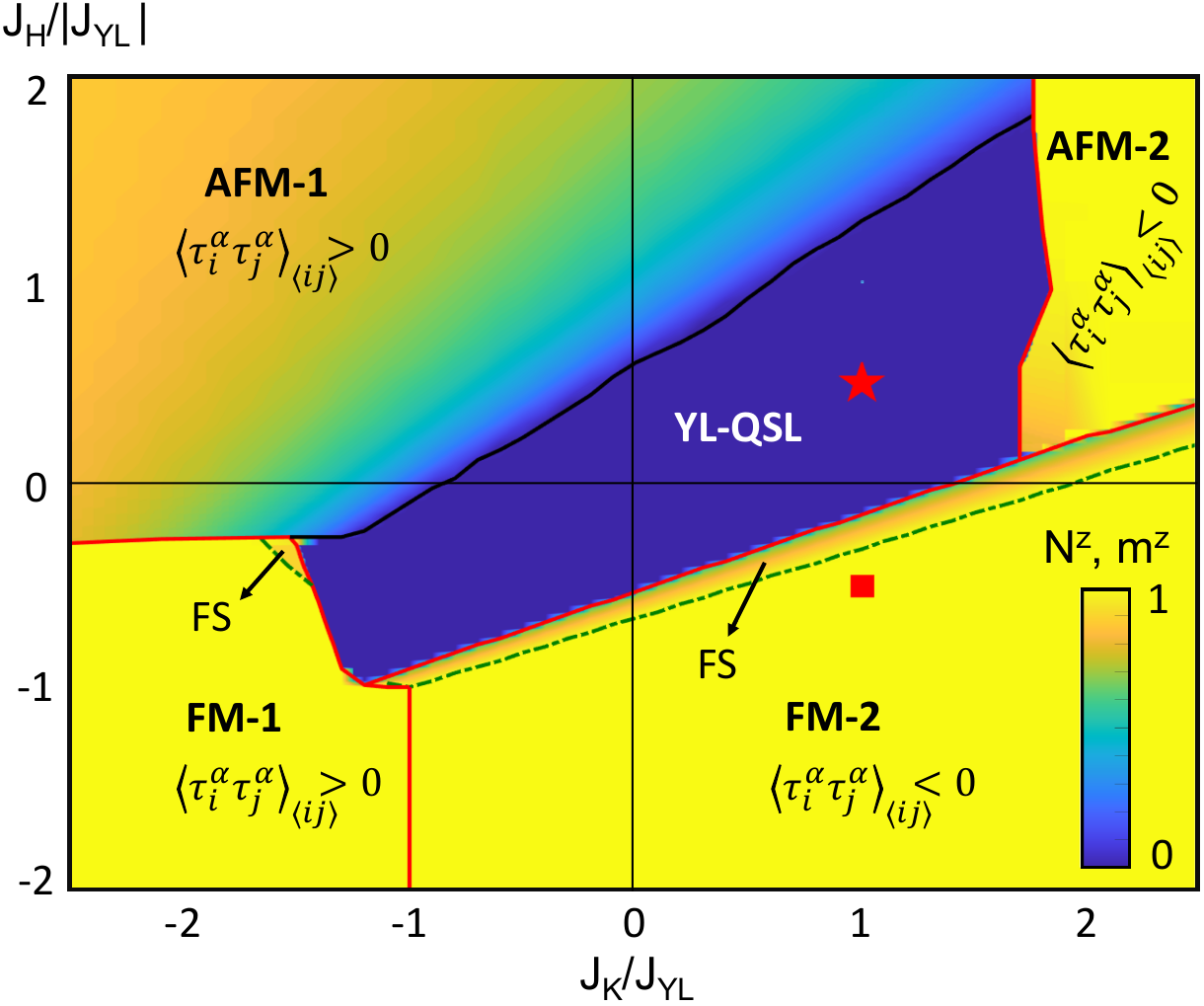}
    \caption{Majorana fermion mean-field phase diagram of the full model as a function of $J_K/Y_{YL}$ and $J_H/|J_{YL}|$. For small Kitaev and Heisenberg couplings, the ground state is a YL spin-orbital liquid with three flavors of itinerant Majorana fermions. For larger $J_K$ and $J_H$, the ground state exhibits FM or AFM ordering in the $\sigma$ degrees of freedom. FM, AFM -1 and -2 are distinguished by the nearest neighbor $\langle \tau_i^\alpha \tau_j^\alpha \rangle_{\langle ij \rangle}$ correlations and are separated by first order transitions. The color plot denotes the magnitude of magnetization and Ne\'el order parameters. The red star and square symbols correspond to the parameters of a microscopic calculation for the realization of YL model\cite{Churchill_npjQM2025}.}
    \label{Fig:3}
\end{figure}

{\it Perturbation theory in the limit of large $J_K$}. As discussed above, in the Kitaev limit the ground state is a Kitaev QSL in $\tau$ sector and a paramagnet in $\sigma$ sector. Therefore, the wave function can be written as a product of the two DOF, $|\Psi\rangle =|\psi_K\rangle_\tau |\prod_i \sigma_i\rangle_\sigma$ where $|\psi_K\rangle_\tau$ is the ground state of the Kitaev model in $\tau$ sector and $\sigma_i = \uparrow$ or $\downarrow$ giving rise to $2^N$ degeneracy. Inclusion of the YL and Heisenberg terms lift this degeneracy through first order perturbation theory

\begin{eqnarray}
    H^{\rm eff} &=& P (H_{YL}+H_H) P \nonumber\\
    &=& \sum_{\langle ij \rangle} (J_H+J_{YL}\langle \psi_K| \tau_i^\alpha \tau_j^\alpha|\psi_K) \rangle)\boldsymbol{\sigma}_i \cdot \boldsymbol{\sigma}_j \nonumber \\
    &=& \sum_{\langle ij \rangle} J^{\rm eff} \boldsymbol{\sigma}_i \cdot \boldsymbol{\sigma}_j
\end{eqnarray}

where $P=|\Psi\rangle \langle \Psi |$ is the projector to the ground state and $J^{\rm eff} = J_H-{\rm sgn}(J_K)0.525 J_{YL}$ as the bond expectation value in the Kitaev model can be calculated exactly\cite{Baskaran_PRL2007}. This implies that infinitesimal Heisenberg or YL couplings can lead to a magnetic order in the $\sigma$ sector, ferromagnetically or antiferromagnetically depending on the sign of $J^{\rm eff}$ while the liquid state in the $\tau$ sector remains unaffected by the perturbation.

{\it Majorana mean field theory}. After establishing the emergence of magnetic order in the limit of large Kitaev coupling, we carry out a self-consistent mean-field analysis of the Hamiltonian to investigate a broader range of parameters. While the YL term involves bilinears of Majorana fermions, the Kitaev and Heisenberg terms have six and four Majorana fermion terms which we decouple in magnetic and non-magnetic channels. Since the model has SU(2) symmetry, we choose the broken symmetry axis for magnetization to be the z-axis, $m_i^z =\langle i c_i^x c_i^y \rangle$. The non-magnetic (Hartree) bond expectation values are $\chi^\alpha = \langle i c_i^\alpha c_j^\alpha \rangle$. The Kitaev term can be decoupled as $-ic_i^x c_i^y c_i^z  c_j^x c_j^y c_j^z \simeq m^z_Am^z_B (ic_i^zc_j^z)+m^z_A\chi^z(ic^x_jc^y_j)+m^z_B\chi^z (ic^x_ic^y_i)+\chi^x\chi^y(ic_i^zc_j^z)+\chi^x\chi^z (ic^y_ic^y_j)+\chi^y\chi^z(ic^x_ic^x_j) -2\chi^z (m^z_A m^z_B+\chi^x\chi^y)$. Similarly the Heisenberg interaction can be decoupled as $(ic^y_ic^z_i)(ic^y_jc^z_j)+(ic^x_ic^z_i)(ic^x_jc^z_j)+(ic^x_ic^y_i)(ic^x_jc^y_j)\simeq m^z_i(ic^x_jc^y_j)+m^z_j(ic^x_ic^y_i)-(\chi^x+\chi^z)(ic^y_ic^y_j)-(\chi^y+\chi^z)(ic^x_ic^x_j)-(\chi^x+\chi^y)(ic^z_ic^z_j)$.
Combining all the terms, the full mean field Hamiltonian reads,

\begin{widetext}
\begin{eqnarray}
    H_{MF}=&&\sum_{\langle ij\rangle}\Big[\big(u_{ij}(J_Y+J_K m^z_im^z_j-J_K\chi^x\chi^y)-J_H(\chi^y+\chi^x)\big)ic_i^zc_j^z+
    \big(u_{ij}(J_Y-J_K\chi^y\chi^z)-J_H(\chi^y+\chi^z)\big)ic_i^xc_j^x+\nonumber \\&& \big(u_{ij}(J_Y-J_K\chi^x\chi^z)-J_H(\chi^x+\chi^z)\big)ic_i^yc_j^y\Big]+ 
\sum_i(J_Ku_{ij}m_j^z\chi^z+J_Hm_j^z)ic_i^xc_i^y+\mathbb{C}.
\label{eq:Hmf}
\end{eqnarray}
\end{widetext}

where the constant energy term is $\mathbb{C}=2J_K(\chi^zm_i^zm^z_j-\chi^x\chi^y\chi^z)+J_H(m^z_im^z_j-\chi^x\chi^y-\chi^y\chi^z-\chi^z\chi^x)$. 
In the remainder, we fix the flux sector to zero-flux and $u_{ij}=1$ for all bonds.

We begin by presenting the phase diagram in the absence of the Heisenberg term ($J_H = 0$), considering only the Kitaev and YL interactions, as shown in Fig.~\ref{Fig:2}. Our self-consistent mean-field analysis reveals that the YL spin-orbital liquid remains stable within the range $-0.85 < J_K/J_{YL} < 1.38$. Beyond this window, the system undergoes magnetic transitions in the $\sigma$ sector, leading to either ferromagnetic (FM) or antiferromagnetic (AFM) order, depending on the sign of $J_K/J_{YL}$. In agreement with perturbative results, AFM order emerges for $J_K/J_{YL} < 0$, while FM order appears for $J_K/J_{YL} > 0$.

AFM transition is second order where the Ne\'el order parameter, $N^z=(m_A^z-m_B^z)/2$, grows continuously where $m^z_{A(B)}$ corresponds to the A(B) sublattice magnetization. The AFM ordering does not affect the excitation spectrum of the $c^z$ Majorana fermions, which remain gapless. However, the $c^x$ and $c^y$ Majorana fermions become gapped. As a result, the low-energy excitation spectrum features a single gapless Majorana flavor with a Dirac-like dispersion.

In contrast, the FM transition for $J_K/J_{YL} > 0$ is first order, characterized by a discontinuous jump in the magnetization. Unlike the AFM phase, the FM order does not gap the $c^x$ and $c^y$ Majorana fermions. Instead, it hybridizes them, leading to a shift in the Dirac cones to higher and lower energies. As a result, the FM phase features a finite Fermi surface (FS) for the $c^x$ and $c^y$ Majorana fermions, while the $c^z$ fermions remain unaffected and retain a Dirac-like spectrum. At $J_K/J_{YL} \simeq 1.91$, a Lifshitz transition occurs where the Majorana Fermi surface disappears, and the system reverts to a single gapless Majorana band with Dirac dispersion. Beyond this point, the FM phase becomes fully polarized, with $m_i^z = 1$, and the non-magnetic mean-field parameters $\chi_i^x$ and $\chi_i^y$ vanish. We emphasize that the difference between the Majorana flavors in the magnetically ordered phases is due to the magnetic ordering along the $z$-axis. In principle, the magnetic order can orient in any direction, as the model possesses SU(2) symmetry in the $\sigma$ degrees of freedom.

We note that the plaquette operator $W_p$ remains conserved for all values of $J_K/J_{YL}$. Although the Kitaev and Yao-Lee models share the same plaquette operator and each independently hosts a spin liquid ground state, a rich phase diagram emerges when both interactions are present. In particular, magnetic order (either FM or AFM) develops in the $\sigma$ sector, while the $\tau$ sector retains its QSL character. This coexistence of magnetic and topological order is commonly referred to as magnetic fragmentation. First predicted in quantum spin ice systems\cite{Brooks_PRX2014, Petit_NatPhys2016, Lefrancois_NatComm2017, Zorko_PRB2019, Mauws_2018,Wang2022}, magnetic fragmentation can arise in a variety of Kitaev spin-orbital models with additional interactions. However, in general the magnetic order is induced by terms such as Heisenberg or Ising interactions\cite{Seifert_PRL2020, Vijayvargia_PRR2023, Vijayvargia_arXiv2025, Fornoville_arXiv2025, Neehus_arXiv2025}. What is particularly striking in our case is that magnetic ordering emerges from the interplay of two terms, each of which independently supports a QSL.

Next, we examine the effects of the Heisenberg interaction on the phase diagram. Fig.~\ref{Fig:3} shows the self-consistent mean-field solutions as a function of $J_K/J_{YL}$ and $J_H/|J_{YL}|$. While the YL spin-orbital liquid is stable for small values of Kitaev and Heisenberg interactions, FM or AFM order in $\sigma$ DOF is stabilized for larger values of $J_K$ and $J_H$. In addition, these magnetically ordered phases are further distinguished by their nearest-neighbor orbital correlations. In FM-1 and AFM-1, the correlations satisfy $\langle \tau_i^\alpha \tau_j^\alpha\rangle_{\langle ij \rangle} > 0$, while in FM-2 and AFM-2, they satisfy $\langle \tau_i^\alpha \tau_j^\alpha\rangle_{\langle ij \rangle} < 0$. Transitions between these phases are first order, except for the AFM-1 to YL-QSL transition, which is second order. FM with a FS persists in FM-2 and a line of Lifshitz transitions connect it to the fully-polarized FM phase. A small region of FM with FS also appears within FM-1, near the boundary between AFM-1 and the YL-QSL phase. It is worth mentioning that all magnetically ordered phases still exhibit topological order in $\tau$ DOF. We note that YL model with AFM Heisenberg interaction has been studied by Ref.~\citenum{Seifert_PRL2020} and our results are in agreement with theirs in the $J_K=0$ limit. 

Naturally, $J_H<0 ~(J_H>0)$ promotes FM (AFM) order. Therefore, $J_K/J_{YL}>0$ and $J_H<0$ quadrant is predominantly ferromagnetic and $J_K/J_{YL}<0$ and $J_H>0$ quadrant is largely antiferromagnetic. In contrast, for the other two quadrants, the Kitaev and Heisenberg terms give rise to frustrated magnetic interactions. This competition favors the YL-QSL and an enhanced region of YL-QSL is stabilized for $J_K/J_{YL},~J_H>0$ and $J_K/J_{YL},~J_H<0$ quadrants.

In the limit of large $|J_K/J_{YL}|$, the mean-field FM–AFM phase boundary agrees closely with the results from perturbation theory. Specifically, for $|J_K/J_{YL}| \gtrsim 10$, the transition occurs at $J_H/J_{YL} = 0.525$ when $J_K > 0$, and at $J_H/J_{YL} = -0.525$ when $J_K < 0$. It is noteworthy that the Majorana mean-field theory yields such accurate results in this regime, despite being formally controlled only when $|J_K/J_{YL}|\ll 1$ and $|J_H/J_{YL}|\ll 1$.

As mentioned earlier, a microscopic superexchange calculation for the realization of the YL model by Ref.~\citenum{Churchill_npjQM2025} involves additional terms. In particular, it takes the form,
\begin{eqnarray}
J(\boldsymbol{\sigma}_i \cdot \boldsymbol{\sigma}_j +1)(2\tau_i^\alpha \tau_j^\alpha - \boldsymbol{\tau}_i \cdot \boldsymbol{\tau}_j  +1)
\end{eqnarray}
Note that we have switched the $\sigma$ and $\tau$ DOF with respect to the convention in Ref.~\citenum{Churchill_npjQM2025}. The $(\boldsymbol{\tau}_i \cdot \boldsymbol{\tau}_j)$ term does not commute with the plaquette operator and therefore it is beyond our scope. The remaining terms can be written as a sum of Kitaev, YL and Heisenberg interactions. In particular they would correspond to $J_K/J_{YL}=1$, $J_H/|J_{YL}|=1/2$ for $J>0$ and $J_K/J_{YL}=1$, $J_H/|J_{YL}|=-1/2$ for $J_{KK}<0$. These two points are marked by red star and square symbols in Fig.~\ref{Fig:3}. Our mean-field analysis implies that the ground state lies in the YL-QSL for $J>0$ and FM-2 for $J<0$. 

Before concluding, we briefly comment on the role of the $(\boldsymbol{\tau}_i \cdot \boldsymbol{\tau}_j)$ and $(\boldsymbol{\sigma}_i \cdot \boldsymbol{\sigma}_j)(\boldsymbol{\tau}_i \cdot \boldsymbol{\tau}_j)$ terms in the phase diagram. These terms break the flux conservation, but the spin liquid phase remains stable for small couplings since the flux excitations are gapped. Moreover, the flux gap in the YL-QSL is three times larger than in the Kitaev model, due to the presence of three flavors of itinerant Majorana fermions. Consequently, we expect the YL-QSL to be more stable than the FM and AFM phases, whose vison gaps are the same as the Kitaev model. In addition, the AFM Kugel-Khomskii model given by $(\boldsymbol{\sigma}_i \cdot \boldsymbol{\sigma}_j +1)(\boldsymbol{\tau}_i \cdot \boldsymbol{\tau}_j+1)$ is known to host a SU(4) spin-orbital liquid\cite{Corboz2012PRX}. Its transition from a YL-QSL is an interesting direction for future study.

In summary, we studied the phase diagram of Yao-Lee model with additional Kitaev and Heisenberg interactions via perturbation and mean-field theory. Even though YL and Kitaev models both have spin liquid ground state by themselves and share the same plaquette operator, the ground state in the presence of both terms can exhibit magnetic order in $\sigma$ DOF while retaining a liquid character in $\tau$ DOF. We showed that inclusion of additional Heisenberg interactions can enhance the magnetic order or suppress it and enlarge the stability of the YL-QSL, depending on its sign. Interesting future directions include exploring the effects of additional interaction terms that break flux conservation, effects of magnetic field, and effects of pressure that alter the orbital composition.

AV and OE acknowledge support from NSF
Award No. DMR-2234352. HYK acknowledges support from NSERC Discovery Grant No. 2022-04601 and the Canada Research
Chairs Program. We thank the ASU Research Computing Center for high performance computing resources. 

MA and AV have contributed equally to this work.

\bibliography{references.bib}
\end{document}


\title{Supplementary material for `Magnetically ordered yet topologically robust phases emerging in concurrent Kitaev spin liquids'}

\author{Muhammad Akram$^{1,2}$\textsuperscript{*}}
\author{Aayush Vijayvargia$^1$\textsuperscript{*}}
\author{Hae-Young Kee$^{3,4}$}
\author{Onur Erten$^1$}

\affiliation{$^1$Department of Physics, Arizona State University, Tempe, AZ 85287, USA}
\affiliation{$^2$Department of Physics, Balochistan University of Information Technology, Engineering and Management Sciences (BUITEMS), Quetta 87300, Pakistan}
\affiliation{$^3$Department of Physics, University of Toronto, Toronto, Ontario, Canada M5S 1A7}
\affiliation{$^4$Canadian Institute for Advanced Research, CIFAR Program in Quantum Materials, Toronto, Ontario, Canada, M5G 1M1}

\maketitle
\section{Alternate Majorana description of $H_K$}
The Majorana fermion description of $H_K$ for the four-dimensional local Hilbert space leads to a term that is not bilinear in Majoranas (other than the bond operators). This results from a Gamma matrix representation and is surprising. Here we circumvent the Gamma matrix representation by introducing separate Majorana descriptions for the spin and orbital DOF and obtain the same final form. Using this more traditional Majorana description will provide an alternate perspective. We restrict ourselves to the $J_H=0$ limit for clarity.

For the spin sector, we have $\sigma^\alpha=ic^0c^\alpha$, and the orbital sector, $\tau^\alpha=id^0d^\alpha$. We use the identity $\sigma^x\sigma^y\sigma^z=\tau^x\tau^y\tau^z=i$ to note that $c^0=ic^xc^yc^z$ and $d^0=id^xd^yd^z$  where $c^0$ and $d^0$ can be thought of as ancillary Majorana fermions.
Using this, one can write: $\vec\sigma=-i\vec c\times \vec c$ and $\vec \tau = -i \vec d \times \vec d$.
Consequently,
$\sigma^\alpha \tau^\beta =-iDc^\alpha d^\beta$, where $D$ is the physical projection operator $D=-ic^0d^0=ic^xc^yc^zd^xd^yd^z$.
In the $D=1$ projected space, these equations can be expressed as, 
\begin{align}
    H_K&=J_K\tau^\alpha_i\tau^\alpha_j=J_Kd^0_id^\alpha_id^0_jd^\alpha_j\nonumber\\
    &=-J_K\text{i}u_{ij}d^0_id^0_j=J_K\text{i}u_{ij}c^0_ic^0_j\nonumber\\
    &=-J_K\text{i}u_{ij}c^x_ic^y_ic^z_ic^x_jc^y_jc^z_j
\end{align}
which matches with eq. (7) in the main text.